\documentstyle[12pt,fleqn]{article}
\def\la{\langle}
\def\ra{\rangle}
\def\beeq{\begin{equation}}
\def\eneq{\end{equation}}
\def\beeqa{\begin{eqnarray}}
\def\eneqa{\end{eqnarray}}

\addtocounter{section}{-1}
\begin{document}

\begin{center}

\vspace{2cm}

{\large {\bf {Coupled exciton model with off-diagonal disorder\\ 
for optical excitations in extended dendrimers
} } }

\vspace{1cm}



{\rm Kikuo Harigaya\footnote[1]{E-mail address: 
\verb+harigaya@etl.go.jp+; URL: 
\verb+http://www.etl.go.jp/+\~{}\verb+harigaya/+}}

\vspace{1cm}

{\sl Physical Science Division,
Electrotechnical Laboratory,\\ 
Umezono 1-1-4, Tsukuba 305-8568, 
Japan}\footnote[2]{Corresponding address}\\
{\sl National Institute of Materials and Chemical Research,\\ 
Higashi 1-1, Tsukuba 305-8565, Japan}\\
{\sl Kanazawa Institute of Technology,\\
Ohgigaoka 7-1, Nonoichi 921-8501, Japan}

\vspace{1cm}

(Received~~~~~~~~~~~~~~~~~~~~~~~~~~~~~~~~~~~)
\end{center}

\vspace{1cm}

\noindent
{\bf Abstract}\\
A phenomenological coupled exciton model is proposed in order 
to characterize optical excitations in extended dendrimers.  
An onsite exciton state is assigned at each phenyl rings 
and a nearest neighbor hopping integral which obeys the 
Gaussian distribution is considered between the exciton states.  
The decreasing optical excitation energy with respect to 
the dendrimer size indicates the presence of exciton funnels 
along the backbone of the dendrimers.  Therefore, the extended 
dendrimers can work as artificial fractal antenna systems 
which capture energy of light.


\pagebreak

Recently, the dendrimer supermolecules with antenna structures
have been investigated extensively.  After capture of light
at the outer edges of the molecule, generated excitons
migrate along the legs of the molecular structures and carry
energy obtained from light.  Then, the excitons move to a certain 
core to localize there, or move to the center of the supermolecule
in order to emit light by recombination of electrons and holes.
Because energy of light is captured and the energy is transfered
by excitons, the supermolecules might act as artificial
molecular antenna.  Therefore, the design of the molecular
structures and their optical properties are quite attractive
in view of scientific interests as well as their potential for
technological applications.

One example of antenna supermolecules has dendrimeric 
structures.  It is a family of molecules composed of 
phenyl rings and acetylene units, namely, extended dendrimers 
[1-5].  Their geometrical structures are illustrated in Fig. 1.
Figure 1 (a) shows the diphenylacetylene with
its abbreviated notation.  Figure 1 (b) shows the family of
extended dendrimers: D4, D10, D25, D58, and D127.  The number
in their names means the number of phenyl rings in the molecules.
In the D4 and D10 dendrimers, each leg is composed of two 
single bonds and one triple bond.  In D25, three central 
legs are made of one phenyl rings and two short legs (shown as 
Fig. 1 (a)).  In D58, three central legs are composed of two 
phenyl rings and three short legs (Fig. 1 (a)).  The next 
connecting legs via the phenyl vertex are like the central legs 
of D25.  In D127, the three central legs are made of three phenyls
and four short legs (Fig. 1 (a)).  In this way, the extended
dendrimers have fractal molecular structures, and the length
of the central legs becomes longer as the size of the dendrimer
becomes larger.

Optical experiments of the extended dendimers [3-5]
show that the optical gap decreases as the dendrimer 
size becomes larger.  The optical spectra have features
which can be understood as contributions from legs
of phenylacetylene oligomers.  The energy of each
feature agrees with that of oligomers.  The excitation
energy becomes smaller as the length of the legs
becomes larger.  Therefore, the presence of exciton
migration pathways along the legs or the backbone of the
supermolecules has been concluded.

Another example of the antenna supermolecules has 
structures with four zinc-containing tetraarylporphylins
linked to a central, unmetallated porphylin through ethyne
bonds [6].  They show migration characters of excitons,
and are interesting as light harvesting antennas.
Such alternative molecular designs, replacement
of molecules at the vertexes, different structure
of leg chains, and so on, have been investigated intensively.
Although various molecular structures are of interests,
we would like to concentrate upon optical excitation
properties of the extended dendrimers in this paper
because of the varieties of possible structures which
will require much more theoretical efforts in future.

Theory of optical excitations in the dendrimers has not
been reported so often, and for example energy transfer 
has been investigated by solving phenomenological 
probability process equations [7].  The rates of the
excitation flow along the legs of the dendrimer have
been introduced, and the mean passage time has been
obtained theoretically.

In this paper, we would like to give rise to a new theoretical
model composed of coupled exciton states with off-diagonal disorder.
We note that the theoretical model of dipole moments [8] 
has been used in order to characterize optical excitations 
of the photosynthetic unit of purple bacteria, which is the 
biological analog of the dendrimeric supermolecule.  In the present 
model, an onsite exciton state is assigned at each phenyl ring.  
There are two possibilities for the nearest neighbor interactions:
(1) When the interactions occur by dipole-dipole couplings,
the direction of the transition dipole moment is by
no means parallel with the electric field of light and
cannot be spatially correlated.  Interaction strengths 
between neighboring dipole moments may vary among positions 
of dipole pairs.  They can be looked as randomly distributed.
Therefore, we assume that the nearest neighbor interactions
obey the Gaussian distribution function.  (2) The second candidate
of the interaction between neighboring exciton states is
an exciton flow characterized with the strengths 
$t_e t_h / \Delta_{\rm ex}$ by perturbation, where $t_e$ and 
$t_h$ are hopping integrals of electrons and holes, 
$\Delta_{\rm ex}$ is the excitation energy of the electron 
hole pair.  We assume that the mean value of the interaction 
is zero and the standard deviation of the interaction $J$ is 
one of theoretical parameters in the Gaussian distribution.
Another theoretical parameter in the model is the site 
energy $E$ which specifies the central energy position of 
excitons in the optical spectra.  The following is our 
tight binding model:
\beeq
H = E \sum_i | i \ra \la i | 
+ \sum_{\la i,j \ra} J_{i,j} ( | i \ra \la j | + {\rm h.c.} ),
\eneq
where $i$ means the $i$th site of the phenyl ring, 
$| i \ra$ is an exciton state at the site $i$,
the sum with $\la i,j \ra$ is taken over neighboring
pairs of sites, and the distribution of $J_{i,j}$ is 
determined by the Gaussian function,
\beeq
P(J_{i,j})=\frac{1}{\sqrt{4\pi}J}
{\rm exp}[-\frac{1}{2} (\frac{J_{i,j}}{J})^2].  
\eneq
Though detailed theoretical treatments are not the same 
altogether, the related exciton models have been used for 
the photosynthetic unit of purple bacteria [8] and the 
J-aggregate systems [9-11].  The diagonalization of eq. (1) 
gives energies of one exciton states measured from the 
energy of the ground state.

The model eq. (1) is diagonalized numerically for the five 
types of the extended dendrimers: D4, D10, D25, D58, and D127.
The lowest eigenvalue always gives the energy position
of the optical absorption edge because the state
with the lowest energy is always allowed for dipole 
transition from the ground state.   This is checked
by looking at the parity of the wave function for each
dendrimer.   In TABLE I, we show the energy of the
absorption edge as a function of the parameters $E$ and $J$.
Here, the number of disorder samples is 10000, and this 
gives the well converged average value of the optical
excitation energy.

Figure 2 shows one example of the comparison of the 
calculation with experiments for the parameters 
$E = 37200{\rm cm}^{-1}$ and $J = 3552{\rm cm}^{-1}$.  
We find fairly good agreement between the experiments and 
calculations.  Two results have the trend that the lowest 
optical excitation energy decreases as the dendrimer 
size becomes larger.  Therefore, it is clarified that the 
presence of exciton migration funnels is well described 
by the present coupled exciton model with off-diagonal 
disorder.  The interaction strength $J$ is one order of 
magnitudes larger than that of the purple bacteria [8], 
indicating the stronger contact between neighboring exciton
states.  When the flow of excitons occur by hopping of 
electrons and holes, the interaction strength is characterized
as $t_e t_h / \Delta_{\rm ex}$ by perturbation.  With assuming
$t_e \sim t_h \sim 0.5t$ and $\Delta_{\rm ex} \sim 2t$ where $t$ 
is the resonance integral of the $\pi$ orbitals of the
phenyl ring, we obtain a characteristic magnitude:
$|J| \sim 0.1 t$.  Here, the values of $t_e$ and $t_h$ 
smaller than $t$ are assumed, because there are three bondings 
between the neighboring phenyls.  We have used $\Delta_{\rm ex} = 2t$, 
as the energy difference between the highest occupied state
and the lowest unoccupied state of a single phenyl is $2t$.
The quantity $|J| \sim 0.1 t$ is also of the same order of 
magnitudes with the above parameter $J = 3552{\rm cm}^{-1}$, 
because $t \sim 2{\rm eV} \sim 25000{\rm cm}^{-1}$.  Thus, our 
theoretical parameter can characterize exciton flows along 
the legs of dendrimers very well.

The extended dendrimers are the rare example where 
$\pi$-conjugated electron systems are present along
the acetylene based legs.  In most of dendrimers 
(see the recent review [12] for example), the systems 
are composed of $\sigma$-bondings rather than $\pi$-bonds.
Owing to the presence of exciton funnels composed of 
$\pi$-bonds, the extended dendrimers can work as  
an artificial fractal antenna which captures
energy of light.

In the compact dendrimers (the another form of 
diphenylacetylene based dendrimers), the optical absorption 
edge less depends on the molecule size [3].  This might come 
from the very huge steric repulsions among neighboring legs, 
and therefore the mutual interactions between legs are hindered 
easily by geometric effects.  The nearly constant absorption 
edge with respect to the dendrimer size indicates that values 
of the interaction strengths $J_{i,j}$ of the compact dendrimers 
are much smaller than those of the extended dendimers when 
the present coupled exciton model is applied to.  Such the 
difference of the parameter values can characterize the 
different dependence on the system size of the extended 
and compact dendimers.

In summary, we have proposed the coupled exciton model
with off-diagonal disorder in order to characterize 
optical excitations in the extended dendrimers.  The 
decreasing optical excitation energy with respect to 
the dendrimer size indicates the presence of exciton 
pathways along the backbone of the dendrimers.

\mbox{}

\begin{flushleft}
{\bf Acknowledgements}
\end{flushleft}

\noindent
Useful discussion with the members of Condensed Matter
Theory Group\\
(\verb+http://www.etl.go.jp/+\~{}\verb+theory/+),
Electrotechnical Laboratory is acknowledged.  
Numerical calculations have been performed on the DEC 
AlphaServer of Research Information Processing System 
Center (RIPS), Agency of Industrial Science and 
Technology (AIST), Japan.

\pagebreak
\begin{flushleft}
{\bf References}
\end{flushleft}

\noindent
$[1]$ Z. Xu and J. S. Moore, Acta Polym. {\bf 45},
83 (1994).\\
$[2]$ C. Devadoss, P. Bharathi, and J. S. Moore,
J. Am. Chem. Soc. {\bf 118}, 9635 (1996).\\
$[3]$ R. Kopelman, M. Shortreed, Z. Y. Shi, W. Tan,
Z. Xu, J. S. Moore, A. Bar-Haim, and J. Klafter,
Phys. Rev. Lett. {\bf 78}, 1239 (1997).\\
$[4]$ M. R. Shortreed, S. F. Swallen, Z. Y. Shi,
W. Tan, Z. Xu, C. Devadoss, J. S. Moore, and R. Kopelman,
J. Phys. Chem. {\bf 101}, 6318 (1997).\\
$[5]$ S. F. Swallen, Z. Y. Shi, W. Tan, Z. Xu,
J. S. Moore, and R. Kopelman, J. Lumin. {\bf 76\&77},
193 (1998).\\
$[6]$ R. W. Wagner, T. E. Johnson, and J. S. Lindsey,
J. Am. Chem. Soc. {\bf 118}, 11166 (1996).\\
$[7]$ A. Bar-Haim and J. S. Klafter, J. Lumin.
{\bf 76\&77}, 197 (1998).\\
$[8]$ T. Ritz, X. Hu, A. Damjanovi\'{c}, 
and K. Schulten, J. Lumin. {\bf 76\&77},
310 (1998).\\
$[9]$ F. C. Spano, J. R. Kuklinski, and S. Mukamel,
Phys. Rev. Lett. {\bf 65}, 211 (1990).\\
$[10]$ F. C. Spano, J. R. Kuklinski, and S. Mukamel,
J. Chem. Phys. {\bf 94}, 7534 (1991).\\
$[11]$ T. Kato, F. Sasaki, S. Abe, and S. Kobayashi,
Chem. Phys. {\bf 230}, 209 (1998).\\
$[12]$ ``Dendrimers", ed. F. V\"{o}gtle
(Springer, Berlin, 1998).\\

\pagebreak

\noindent
TABLE I.  The energy of the lowest optical excitation.\\
\\
\begin{tabular}{lc} \hline \hline
Dendrimer &   Absorption edge\\ \hline
D4    & $E - 1.6037J$ \\
D10   & $E - 2.3346J$ \\
D25   & $E - 2.8078J$ \\
D58   & $E - 3.1588J$ \\
D127  & $E - 3.4338J$ \\ \hline \hline
\end{tabular}

\mbox{}

\mbox{}

\begin{flushleft}
{\bf Figure Captions}
\end{flushleft}

\mbox{}

\noindent
Fig. 1.  (a) Diphenylacetylene and its abbreviated notation.
The closed circle means a phenyl ring, and the connecting line means
two single bonds and one triple bond.  (b)  The extended dendimers:
D4, D10, D25, D58, and D127.  The number in their names is
number of phenyl rings in fractal supermolecules.

\mbox{}

\noindent
Fig. 2. Optical absorption edge of the five dendrimers:
D4, D10, D25, D58, and D127.  The crosses are taken from
experiments [3], and the squares are the calculated results
with $E=37200{\rm cm}^{-1}$ and $J=3552{\rm cm}^{-1}$.

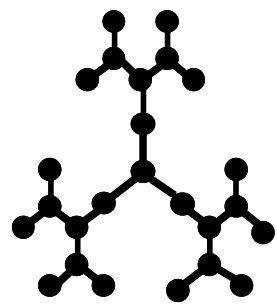
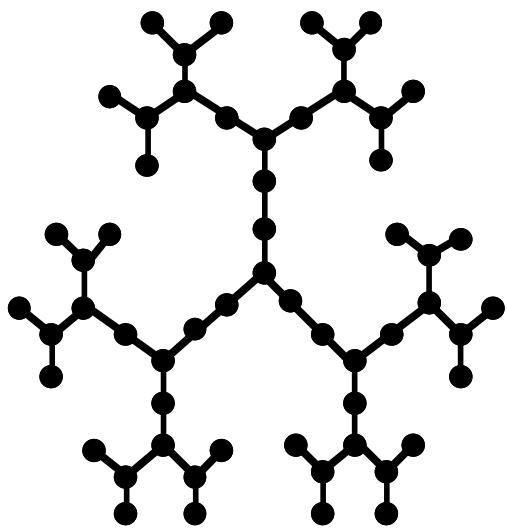
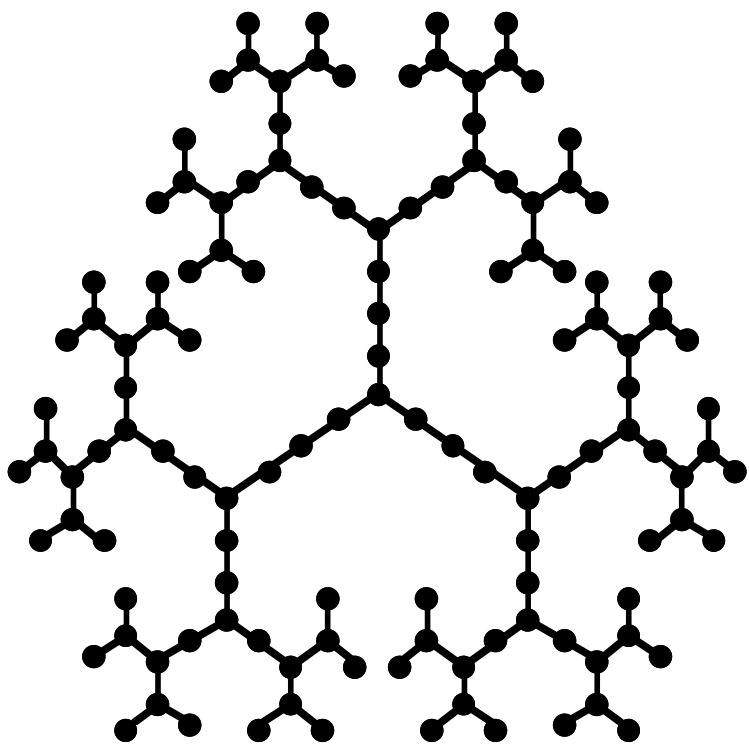
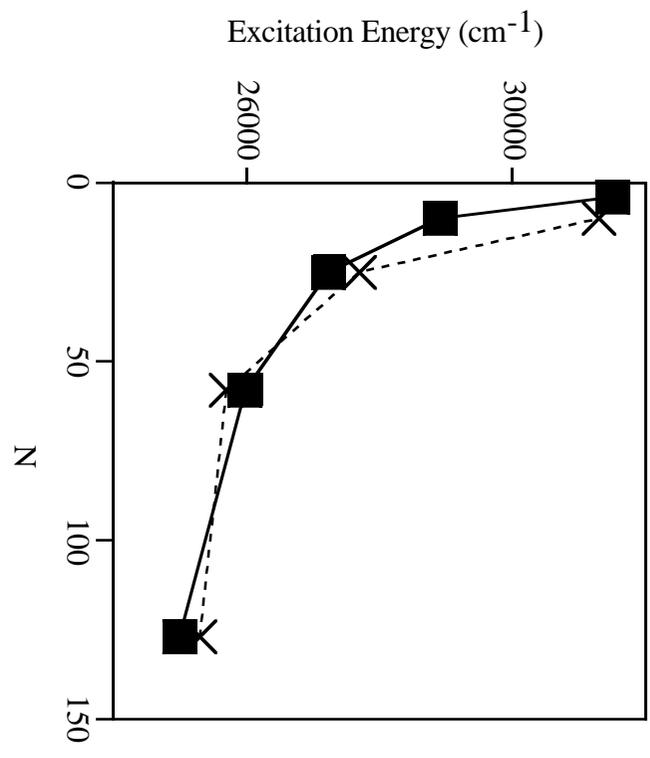
\end{document}